\def\AAPREPRINT{}
\documentclass{aa}

\makeatletter
\renewcommand*\aa@acknowledgementsfont{\small}
\renewcommand*\aa@bibliographyfont{\small}
\makeatother

\usepackage[T1]{fontenc}
\usepackage[varg]{txfonts}
\usepackage{graphicx}
\usepackage{natbib}
\bibpunct{(}{)}{;}{a}{}{,}
\usepackage{stfloats}
\newcommand{\nextpanel}{\hfil\hfil}
\newcommand{\panelrow}[2]{\centerline{\hbox to #1{\hfil #2\hfil}}}

\graphicspath{{figures/}{./}}

\begin{document}

\title{Physical assumptions versus nature: the shape of the heliosphere}

\author{Romana Ratkiewicz\inst{1}\corrauth{roma@cbk.waw.pl}}

\institute{Space Research Centre, Polish Academy of Sciences,
           Bartycka 18A, 00-716 Warsaw, Poland}

\date{Received ... / Accepted ...}

\abstract
{The Interstellar Probe study describes a mission concept with a launch
window beginning in 2036, aiming to reach several hundred astronomical
units beyond the heliopause within a time frame of at least 50 years. One
of the fundamental questions related to this mission is where the
boundaries of the heliosphere are and how thick they are. To answer it we
need to know the shape of the heliosphere. Over the past 60 years numerous
models of the heliosphere have been developed. For a long time these models
depicted a heliosphere resembling a comet. Since around 2015 models have
been presented that differ significantly from the shape of a comet.}
{The aim of this paper is to emphasize that the obtained image of the
heliosphere depends on the model assumptions contained in these articles.}
{We review numerous publications examining the shape of the heliosphere.
All of these articles use a magnetohydrodynamic (MHD) model of the
heliosphere together with various assumptions that simplify the physics of
the interaction of the solar wind flowing from the Sun with the surrounding
interstellar medium.}
{The result of all these studies is a shape of the heliosphere described as
either croissant-like or cometary-like, depending on the assumptions made.}
{We are still uncertain which model represents the real heliosphere. Nature
requires a theoretical model that is more accurate. In our opinion the
interaction of the solar wind with the interstellar medium must be
described by the Boltzmann system of equations using artificial
intelligence. In the context of the planned space mission this conclusion
is crucial.}


\titlerunning{Physical assumptions versus nature}
\authorrunning{R. Ratkiewicz}

\maketitle
\ifdefined\AAPREPRINT\nolinenumbers\fi

\section{Introduction}

\begin{figure}
  \centering
  \includegraphics[width=\hsize]{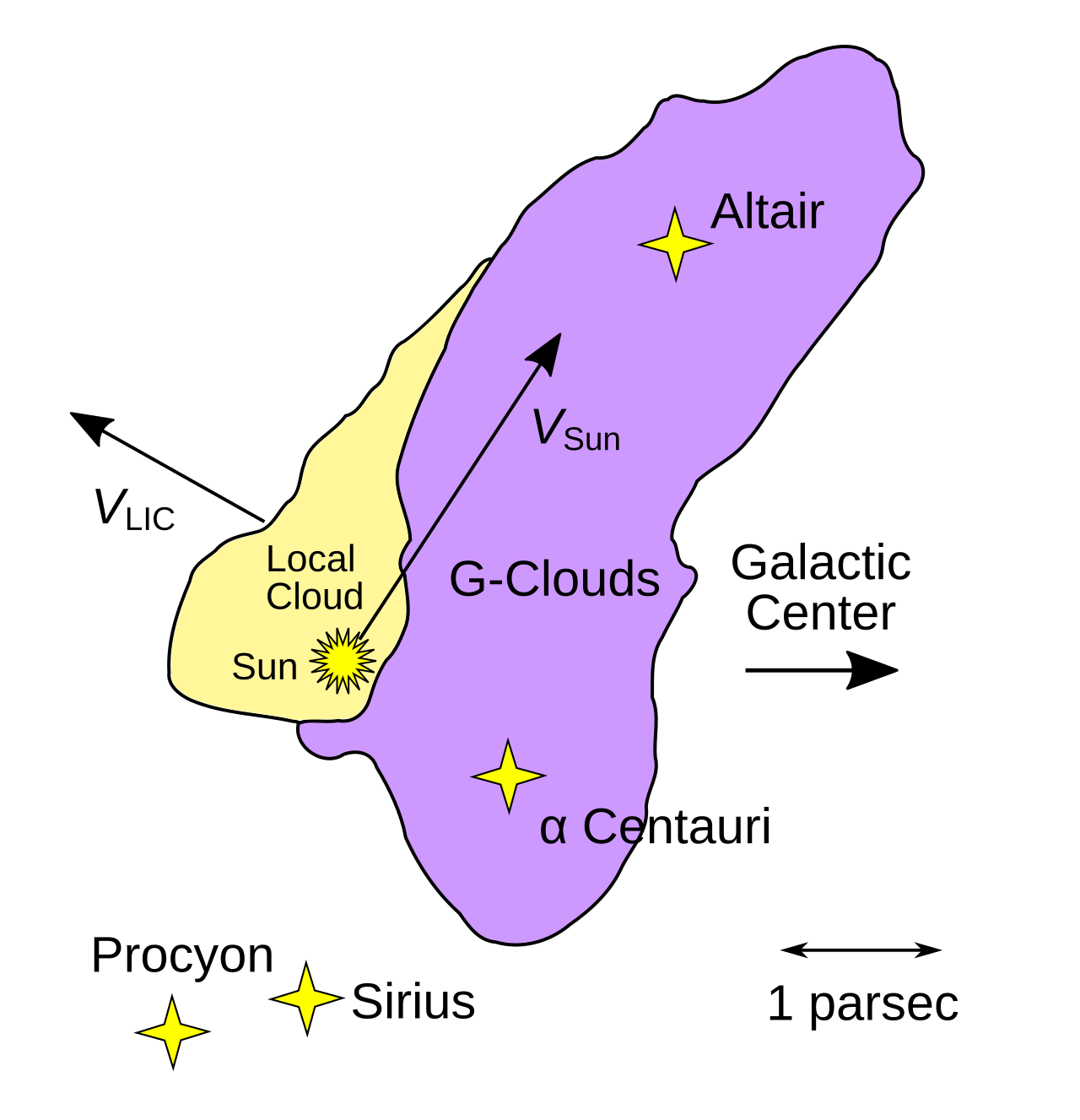}
  \caption{Map showing the Sun near the edge of the Local Interstellar
    Cloud (LIC) and Alpha Centauri about 4 light-years from the Sun in the
    neighbouring G-Cloud complex. Below it is Sirius, the brightest star in
    Canis Major. Diagram after a NASA original, vectorised by
    B.~Jankuloski; Wikimedia Commons, CC0.}
  \label{fig:lic}
\end{figure}

The Sun is one of a hundred billion stars and astrospheres in the Galaxy
that are bursting through a vast interstellar medium made up of supernova
remnants and condensed stellar outbursts \citep{opher_loeb2022}. About
60\,000 years ago the Sun entered the Local Interstellar Cloud (LIC). For
the past few thousand years the Sun appears to have been traversing the
outer layer of the LIC and is now at its very edge (Fig.~\ref{fig:lic}).

Our star is surrounded by the heliosphere, a cavity carved out of the Local
Interstellar Medium (LISM) by the expanding solar corona. We owe this
knowledge to Eugene Parker, who discovered that the solar corona must be
expanding and named this coronal expansion the solar wind (SW). The
resultant of the velocity vectors of interstellar matter and of the Sun
moving through our Galaxy shows the LISM inflow toward the Sun. Parker
created the first model of the interaction of the solar wind plasma with
the interstellar matter plasma using the equations of hydrodynamics
\citep{parker1958}. The interface created as a result of this interaction,
being the surface of pressure equilibrium of both media and called the
heliopause (HP), had a shape similar to the shape of a comet moving near
the Sun \citep{parker1963}. Since the first model by Parker, many models
suggested a comet-like shape
\citep{holzer1989, baranov1991, baranov1993, pauls1996, pauls1997,
washimi1996, pogorelov1998, pogorelov2013, pogorelov2015, ratkiewicz1998,
ratkiewicz2000, ratkiewicz2012, ratkiewicz2006, zank1999, fahr2000,
opher2009, chalov2010, washimi2011, provornikova2014, izmodenov2015,
bladek2023}.

A self-consistent hydrodynamic model of the SW--LISM interaction, which
took into account the influence of the LISM ionised component and of the
LISM neutral hydrogen atoms, was made by \citet{baranov1993}. A five-fluid
hydrodynamic approach to model the SW--LISM interaction was created by
\citet{fahr2000}. The first magnetohydrodynamic (MHD) models of the
heliosphere, in which the interstellar magnetic field vector (ISMF) is
arbitrarily inclined to the LISM velocity vector, were developed by
\citet{pogorelov1998}, published in JGR in January 1998,
\citet{linde1998}, published in JGR in February 1998, and
\citet{ratkiewicz1998}, accepted by A\&A in February 1998.

Complex interactions of plasmas of both media, magnetic fields and neutral
interstellar gas, starting near the Sun and acting throughout the
heliosphere, are responsible for shaping the boundary between the solar
wind and the local interstellar medium. Unfortunately, the boundary region
composed of the outer heliosphere (heliosheath) and the very local
interstellar medium (VLISM) is an almost completely unexplored region of
space \citep{brandt2023}. For this reason the boundary conditions for
interstellar matter, despite experimental data from space missions, are
still poorly known. And none of these numerical models of the heliosphere
meet the requirements of combining all measurement data from space missions
into one numerical solution. This imperfection contributed to further
searches for appropriate models of the heliosphere. From 2015 to 2020 three
models of the heliosphere were created that distinctly differ from the
comet shape: a croissant-like shape based on Voyager~1 data
\citep{opher2015}, a bubble shape based on Cassini data
\citep{dialynas2017}, and a croissant-like and bubble shape
\citep{opher2020}.

Since the heliosphere is formed by the interaction of the SW with the LISM,
the shape and position of the heliospheric boundary depend on the
parameters of the interacting flows, namely (a) solar wind plasma with
interstellar plasma and (b) solar wind plasma with the non-ionised
interstellar medium, and on the processes occurring within these flows,
such as interplanetary (IMF) and interstellar (ISMF) magnetic fields,
ionised and neutral LISM components, energetic neutral atoms (ENA),
pick-up ions (PUI), anomalous cosmic rays (ACR), galactic cosmic rays
(GCR), time-dependent phenomena, latitudinal SW dependence, the effective
tilt of the Sun's magnetic dipole relative to the spin axis, the
heliospheric current sheet (HCS), magnetic reconnection, dissipative
processes, instabilities, turbulence, and others that we have not
discovered yet. Up to now no model of the heliosphere has accounted for all
of these processes. However, many of the processes mentioned above are
partially combined. Depending on them, the shape of the heliosphere is
croissant-like or cometary-like. There is currently an open debate about
whether the heliosphere resembles a comet with a long tail or a croissant
with a shorter one. Since the goal of this paper is to understand how and
what assumptions influence the final picture of the heliosphere, the
following 18 papers were selected to answer these questions:
\citet{opher2015}, \citet{pogorelov2015}, \citet{izmodenov2015},
\citet{zirnstein2017}, \citet{opher2017}, \citet{pogorelov2017},
\citet{izmodenov2020}, \citet{opher2020}, \citet{michael2021},
\citet{opher2021}, \citet{kornbleuth2021}, \citet{michael2022},
\citet{zirnstein2022}, \citet{ma2025a}, \citet{heerikhuisen2025},
\citet{onubogu2025}, \citet{ma2025b}, and \citet{reisenfeld2026}. Among
them, some articles, under certain assumptions, did not provide any of the
above answers, leaving them for further research.

The article is organized as follows: Section~\ref{sec:assumptions} shows
which of these articles describe the shape of the heliosphere as
comet-like or croissant-like, and which ones, with more complex
assumptions, do not always answer the question of the shape of the
heliosphere, deferring this research to future studies.
Section~\ref{sec:examples} provides examples of the discussion of
assumptions by \citet{kornbleuth2021} and \citet{heerikhuisen2025}, based
mainly on \citet{opher2015}, \citet{pogorelov2015} and
\citet{izmodenov2015}. Section~\ref{sec:issues} addresses issues
concerning instabilities in the heliosheath, constraints related to solar
wind speed, the role of turbulence and the IMAP mission.
Section~\ref{sec:conclusions} presents the conclusions.

\section{Assumptions leading to the shape of the heliosphere}
\label{sec:assumptions}

Five of the articles listed below are presented in
Sect.~\ref{sec:examples} in a subsection of their own, and the
cross-reference after such an entry points to it. Part of the others are
not discussed there directly but are cited by the authors of those five
publications, and are marked accordingly. The four entries carrying no
cross-reference at all are presented in Sect.~\ref{sec:issues}.

\subsection{Articles with the comet-like heliosphere shape}

\begin{itemize}
  \item \citet{pogorelov2015} (Sect.~\ref{sec:pogorelov2015})
  \item \citet{izmodenov2015} (Sect.~\ref{sec:izmodenov2015})
  \item \citet{pogorelov2017} (cited in Sect.~\ref{sec:examples})
  \item \citet{izmodenov2020} (cited in Sect.~\ref{sec:examples})
\end{itemize}

\subsection{Articles with the croissant-like heliosphere shape}

\begin{itemize}
  \item \citet{opher2015} (Sect.~\ref{sec:opher2015})
  \item \citet{opher2017} (cited in Sect.~\ref{sec:examples})
  \item \citet{opher2020} (cited in Sect.~\ref{sec:examples})
  \item \citet{michael2021} (cited in Sect.~\ref{sec:examples})
\end{itemize}

\subsection{Articles with more complex assumptions}

\begin{itemize}
  \item \citet{zirnstein2017} (cited in Sect.~\ref{sec:examples})
  \item \citet{opher2021} (cited in Sect.~\ref{sec:examples})
  \item \citet{kornbleuth2021} (Sect.~\ref{sec:kornbleuth2021})
  \item \citet{michael2022} (cited in Sect.~\ref{sec:examples})
  \item \citet{zirnstein2022} (cited in Sect.~\ref{sec:examples})
  \item \citet{heerikhuisen2025} (Sect.~\ref{sec:heerikhuisen2025})
  \item \citet{ma2025a}
  \item \citet{onubogu2025}
  \item \citet{ma2025b}
  \item \citet{reisenfeld2026}
\end{itemize}

\section{Assumptions in selected publications}
\label{sec:examples}

In this section we cite excerpts from \citet{kornbleuth2021} and
\citet{heerikhuisen2025}, in which the authors summarise the crux of the
purpose of our article by demonstrating how the assumptions of a numerical
model influence the shape of the heliosphere. Because these articles are
mainly based on \citet{opher2015}, \citet{izmodenov2015} and
\citet{pogorelov2015}, their abstracts and figures are cited first.

\clearpage
\subsection{Opher et al. (2015)}
\label{sec:opher2015}

This is the first model introducing a shape of the heliosphere
significantly different from the cometary one.

\begin{figure*}[!b]
  \centering
  \panelrow{16cm}{(a)\nextpanel(b)\nextpanel(c)}
  \includegraphics[width=16cm]{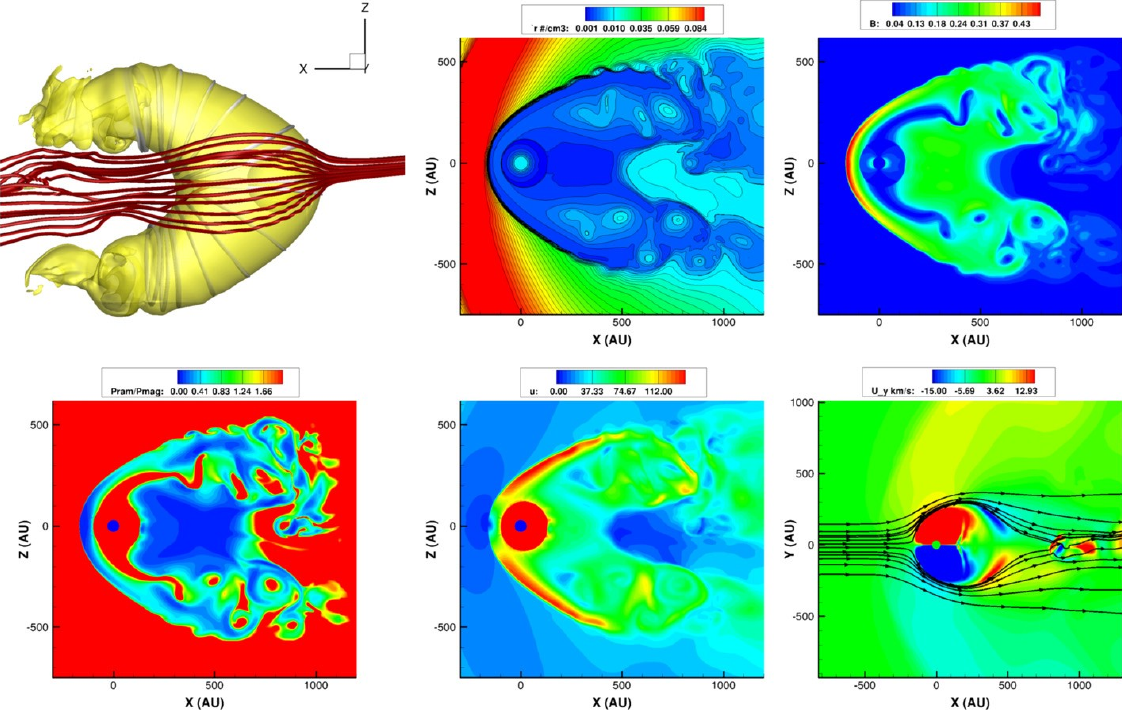}
  \panelrow{16cm}{(d)\nextpanel(e)\nextpanel(f)}
  \caption{Two-lobe structure for the case with no interstellar magnetic
    field (a--f). The panels are shown at the end of the simulation at
    865~yr. (a) View from the west from the point of view of looking from
    the ISM toward the nose of the heliosphere. The grey lines are the
    magnetic field lines and the red the interstellar wind velocity
    streamlines that stream from right to left in panel (a). The yellow
    iso-surface at $\ln T = 12.7$ denotes the heliopause. Panels (b)--(e)
    are in the meridional plane at $y = 0$~AU. Contours are (b) density,
    (c) magnetic field, (d) $P_{\rm ram}/P_{B}$, where $P_{\rm ram}$ is the
    ram pressure and $P_{B}$ is the magnetic pressure, (e) speed, and (f)
    cut in the equatorial plane at $z = 0$~AU; contours are the meridional
    flows $U_{y}$. The black lines are the velocity streamlines. The view
    in panel (a) is reversed from panels (b)--(e). Reproduced from Fig.~1
    of \citet{opher2015}.}
  \label{fig:opher2015noB}

  \vspace{5mm}
  \includegraphics[width=14cm]{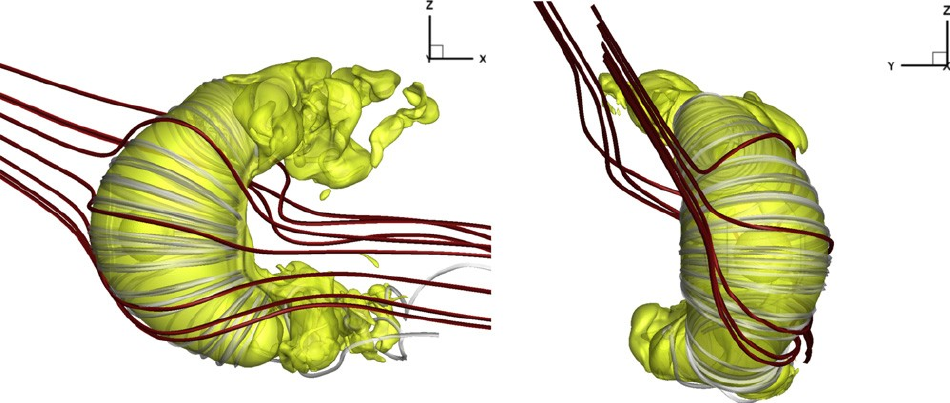}
  \panelrow{14cm}{(a)\nextpanel(b)}
  \caption{Two-lobe structure for the case with an interstellar magnetic
    field. The heliopause is captured at the iso-surface of
    $\ln T = 12.7$; the grey lines are the solar magnetic field lines and
    the red lines are the interstellar magnetic field. (a) Side view;
    (b) nose view. Reproduced from Fig.~2 of \citet{opher2015}.}
  \label{fig:opher2015withB}
\end{figure*}

\begin{quote}
  ``The classic accepted view of the heliosphere is a quiescent, comet-like
  shape aligned in the direction of the Sun's travel through the
  interstellar medium (ISM) extending for thousands of astronomical units
  (AUs). Here, we show, based on magnetohydrodynamic (MHD) simulations,
  that the tension (hoop) force of the twisted magnetic field of the Sun
  confines the solar wind plasma beyond the termination shock and drives
  jets to the north and south very much like astrophysical jets. These jets
  are deflected into the tail region by the motion of the Sun through the
  ISM similar to bent galactic jets moving through the intergalactic
  medium. The interstellar wind blows the two jets into the tail but is not
  strong enough to force the lobes into a single comet-like tail, as
  happens to some astrophysical jets. Instead, the interstellar wind flows
  around the heliosphere and into the equatorial region between the two
  jets. As in some astrophysical jets that are kink unstable, we show here
  that the heliospheric jets are turbulent (due to large-scale MHD
  instabilities and reconnection) and strongly mix the solar wind with the
  ISM beyond 400~AU. The resulting turbulence has important implications
  for particle acceleration in the heliosphere. The two-lobe structure is
  consistent with the energetic neutral atom (ENA) images of the heliotail
  from IBEX where two lobes are visible in the north and south and the
  suggestion from the Cassini ENAs that the heliosphere is `tailless.'\,''
  [Figs.~\ref{fig:opher2015noB} and~\ref{fig:opher2015withB}]
\end{quote}

\subsection{Izmodenov \& Alexashov (2015)}
\label{sec:izmodenov2015}

\begin{figure*}[p]
  \centering
  \panelrow{14cm}{(A)\nextpanel(B)}
  \includegraphics[width=14cm]{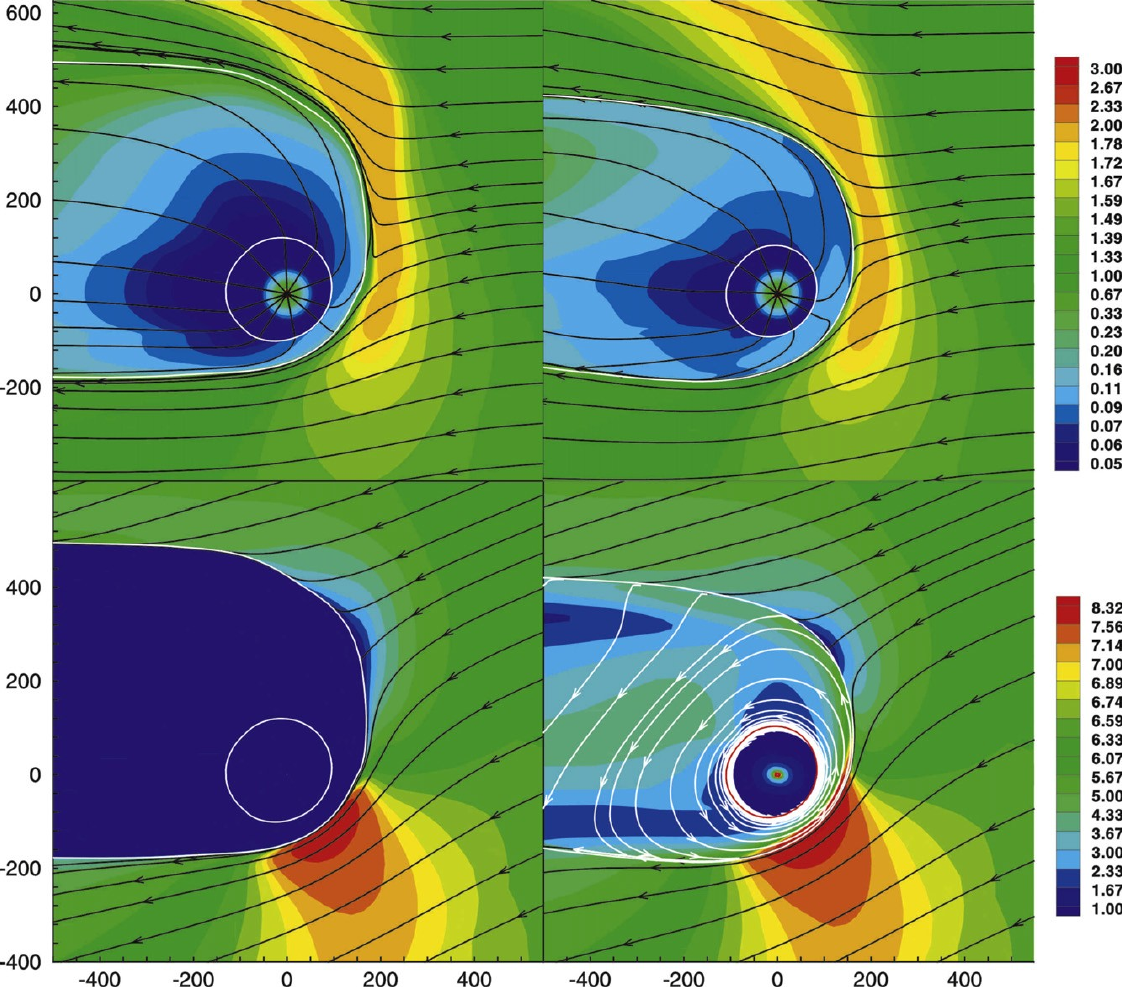}
  \panelrow{14cm}{(C)\nextpanel(D)}
  \caption{Panels (A) and (B): plasma streamlines and isolines of the
    plasma density normalised to the proton density in the LISM. Panels (C)
    and (D): magnetic field lines and isolines of the magnetic field
    magnitude in dimensionless units. Left panels (A) and (C) correspond to
    Model~1 without HMF, right panels (B) and (D) correspond to Model~2
    with HMF. All of the panels are made in the $ZX$ plane determined by
    the interstellar velocity and magnetic field vectors. Reproduced from
    Fig.~3 of \citet{izmodenov2015}.}
  \label{fig:izmodenov2015}
\end{figure*}

\begin{quote}
  ``This paper provides a detailed description of the latest version of our
  model of the solar wind (SW) interaction with the local interstellar
  medium (LISM). This model has already been applied to the analysis of
  Ly$\alpha$ absorption spectra toward nearby stars and for analyses of
  Solar and Heliospheric Observatory/SWAN data. Katushkina et al.\ (this
  issue) used the model results to analyze IBEX-Lo data. At the same time,
  the details of this model have not yet been published. This is a
  three-dimensional (3D) kinetic-magnetohydrodynamical (MHD) model that
  takes into account SW and interstellar plasmas (including $\alpha$
  particles in SW and helium ions in LISM), solar and interstellar magnetic
  fields, and interstellar hydrogen atoms. The latitudinal dependence of SW
  and the actual flow direction of the interstellar gas with respect to the
  Sun are also taken into account in the model. It was very essential that
  our numerical code was developed in such a way that any numerical
  diffusion or reconnection across the heliopause were not allowed in the
  model. The heliospheric current sheet is a rotational discontinuity in
  the ideal MHD and can be treated kinematically. In the paper, we focus in
  particular on the effects of the heliospheric magnetic field and on the
  heliolatitudinal dependence of SW.''
  [Fig.~\ref{fig:izmodenov2015}]
\end{quote}

\subsection{Pogorelov et al. (2015)}
\label{sec:pogorelov2015}

\begin{figure*}[p]
  \centering
  \includegraphics[width=17cm]{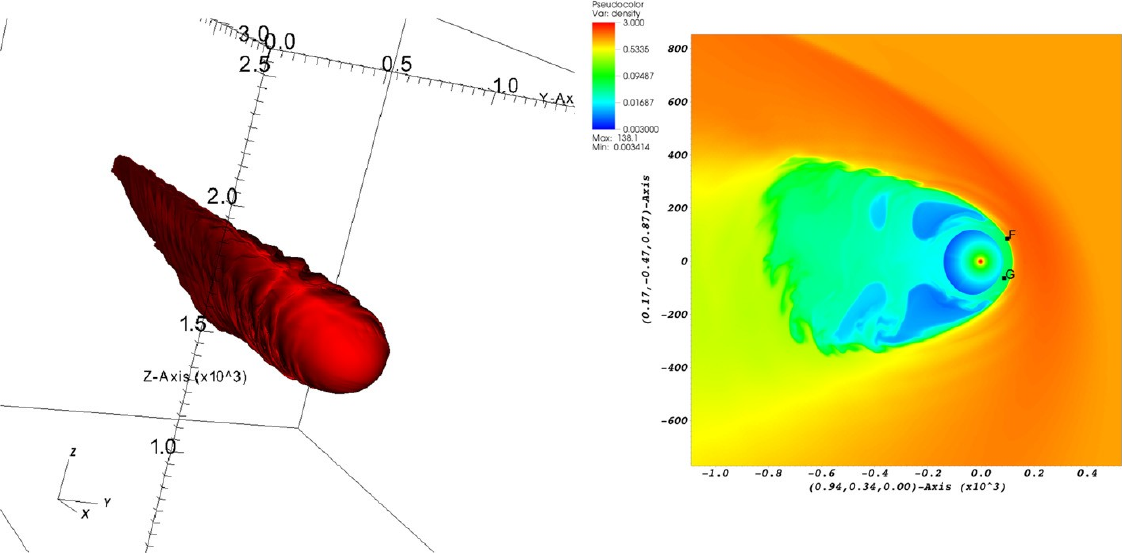}
  \caption{\emph{Left panel:} 3D view of the heliopause showing that the
    heliotail is compressed by the ISMF approximately in the direction
    perpendicular to the $BV$-plane while preserving an unsplit structure.
    \emph{Right panel:} cross-section of the heliosphere by the V1--V2
    trajectory plane (on 2015 July 1). The squares with the attached
    letters, F and G, show the V1 and V2 locations. Reproduced from Fig.~5
    of \citet{pogorelov2015}.}
  \label{fig:pogorelov2015}
\end{figure*}

\begin{quote}
  ``The heliotail is formed when the solar wind (SW) interacts with the local
  interstellar medium (LISM) and is shaped by the interstellar magnetic
  field (ISMF). While there are no spacecraft available to perform in situ
  measurements of the SW plasma and heliospheric magnetic field (HMF) in
  the heliotail, it is of importance for the interpretation of measurements
  of energetic neutral atom fluxes performed by Interstellar Boundary
  Explorer. It has been shown recently that the orientation of the
  heliotail in space and distortions of the unperturbed LISM caused by its
  presence may explain the anisotropy in the TeV cosmic ray flux detected
  in air shower observations. The SW flow in the heliotail is a mystery
  itself because it is strongly affected by charge exchange between the SW
  ions and interstellar neutral atoms. If the angle between the Sun's
  magnetic and rotation axes is constant, the SW in the tail tends to be
  concentrated inside the HMF spirals deflected tailward. However, the
  twisted field soon becomes unstable and the reason for the SW collimation
  within a two-lobe structure vanishes. We demonstrate that kinetic
  treatment of the H atom transport becomes essential in this case for
  explaining the lobe absence further along the tail. We show that the
  heliotail flow is strongly affected by the solar cycle that eliminates
  artifacts, which is typical of solutions based on simplifying
  assumptions. The heliopause in the tail is subject to Kelvin--Helmholtz
  instability, while its orientation and shape are determined by the ISMF
  direction and strength.''
  [Fig.~\ref{fig:pogorelov2015}]
\end{quote}

\subsection{Two shapes from the same boundary conditions}
\label{sec:samebc}

As noted at the beginning of this section, we now consider the question of
whether two different shapes of the heliosphere can be obtained under the
same boundary conditions. A direct answer is provided by
\citet{kornbleuth2021} and \citet{heerikhuisen2025}, in which the authors
compare the Boston University and Moscow University models
\citep{kornbleuth2021} and the Boston University and University of Alabama
in Huntsville models \citep{heerikhuisen2025}, and prove that this is
possible.

\subsubsection{Kornbleuth et al. (2021)}
\label{sec:kornbleuth2021}

A significant citation from this publication is: ``Both models use
identical boundary conditions to compare how different numerical approaches
and physical assumptions contribute to the heliospheric solution.''

From the abstract:

\begin{quote}
  ``Global models of the heliosphere are critical tools used in the
  interpretation of heliospheric observations. There are several
  three-dimensional magnetohydrodynamic (MHD) heliospheric models that rely
  on different strategies and assumptions. Until now only one paper has
  compared global heliosphere models, but without magnetic field effects.
  We compare the results of two different MHD models, the BU and Moscow
  models. Both models use identical boundary conditions to compare how
  different numerical approaches and physical assumptions contribute to the
  heliospheric solution. Based on the different numerical treatments of
  discontinuities, the BU model allows for the presence of magnetic
  reconnection, while the Moscow model does not. Both models predict
  collimation of the solar outflow in the heliosheath by the solar magnetic
  field and produce a split tail where the solar magnetic field confines
  the charged solar particles into distinct north and south columns that
  become lobes. In the BU model, the interstellar medium (ISM) flows
  between the two lobes at large distances due to MHD instabilities and
  reconnection. Reconnection in the BU model at the port flank affects the
  draping of the interstellar magnetic field in the immediate vicinity of
  the heliopause. Different draping in the models cause different ISM
  pressures, yielding different heliosheath thicknesses and boundary
  locations, with the largest effects at high latitudes. The BU model
  heliosheath is 15\% thinner and the heliopause is 7\% more inwards at the
  north pole relative to the Moscow model. These differences in the two
  plasma solutions may manifest themselves in energetic neutral atom
  measurements of the heliosphere.''
\end{quote}

From Sect.~2, Model:

\begin{quote}
  ``The two numerical models of the heliosphere used in this work are the
  Moscow model (Izmodenov \& Alexashov 2015, 2020) and the BU model (Opher
  et al.\ 2015; Michael et al.\ 2021). Both models are 3D kinetic-MHD
  models that treat the cold solar wind plasma and hot pick-up ion (PUI)
  plasma as a single fluid, while treating neutral particles kinetically.
  Both models use the same inner boundary conditions for the solar wind, an
  averaged solar cycle condition from the years 1995 to 2017 and the same
  outer boundary conditions for the interstellar neutral H and protons
  (Izmodenov \& Alexashov 2020).''
\end{quote}

\begin{quote}
  \textbf{Table 2. Summary of conclusions from this work}

  \emph{Question:} How does the size of the heliosphere change between the
  two models?\\
  \emph{Answer:} The heliopause is closer to the Sun at high latitudes in
  the BU model than in the Moscow model, leading to a more compressed
  heliosphere at the poles.

  \emph{Question:} How does the heliotail compare between the two models?\\
  \emph{Answer:} Both models show collimation of the heliosheath plasma by
  the solar magnetic field. They also both produce a split tail since the
  solar magnetic field confines the solar particles into two distinct north
  and south columns that become lobes. In the BU model, reconnection allows
  the ISM to flow between the two lobes at large distances. The
  low-latitude heliotail is shorter in the BU model while the Moscow model
  has a long, comet-like tail.

  \emph{Question:} What assumptions have the largest impact on the
  heliospheric structure?\\
  \emph{Answer:} Magnetic reconnection at the heliopause in the BU model
  leads to an increase in the magnetic pressure outside of the heliosphere,
  which compresses the heliosphere. Non-ideal MHD effects contribute to a
  split tail in the BU model.

  \emph{Question:} How do the plasma properties in the heliosheath
  compare?\\
  \emph{Answer:} Within 100~au of the termination shock, the mass flux is
  similar between the two models, showing an organization by the solar
  magnetic field.

  \emph{Question:} How do the plasma properties in the ISM compare?\\
  \emph{Answer:} The ISM conditions are similar out to roughly 50~au ahead
  of the heliopause, where the interaction between the interstellar and
  solar magnetic fields at the heliopause begins to have an effect on the
  plasma in the BU model.
\end{quote}

From the conclusions:

\begin{quote}
  ``Table~2 summarizes our results. We compared two MHD models using
  identical inner and outer boundary conditions for the protons and neutral
  H atoms, and investigated how the different assumptions present in each
  model influenced the results. Both models produce a split tail since the
  solar magnetic field confines the solar particles into distinct north and
  south columns that become lobes. In the BU model, the ISM flows between
  the two lobes at large distances. MHD instabilities in the BU model
  (Opher et al.\ 2021), which are absent in the Moscow model, affect the
  heliotail and create a short, split tail in contrast with the long tail
  of the Moscow model. The BU model gives a smaller heliosphere with a
  heliopause closer to the Sun than predicted in the Moscow model in all
  directions. The communication between the solar and interstellar magnetic
  fields at the heliopause in the BU model allows for magnetic reconnection
  in the port flank (Opher et al.\ 2017) and leads to a change in the
  magnetic topology, which does not occur in the Moscow model. The BU model
  uses a polarity for the solar magnetic field that suppresses reconnection
  at the nose. Reconnection occurs in the port flank, the main location
  where reconnection should proceed since diamagnetic effects suppress
  reconnection at the nose (Opher et al.\ 2017). Magnetic reconnection at
  the heliopause in the BU model twists the interstellar magnetic field and
  leads to an increase in the interstellar magnetic pressure relative to
  the Moscow model. In comparing the two models, we see a short
  low-latitude tail in the BU model and a long tail in the Moscow model.
  The question of the structure of the heliotail is currently an active and
  vigorous topic of debate in the community between several groups. This
  paper establishes that both the BU and Moscow models predict magnetic
  confinement in the heliotail. This supports the earlier conclusions of
  the split arrangement of the heliotail made by Opher et al.\ (2015),
  Izmodenov \& Alexashov (2015), and Pogorelov et al.\ (2015), which had
  still been debated. The main difference between the two models is the ISM
  piercing the two lobes in the BU model, which is absent in the Moscow
  model. The next key step is to pin down the driver for the splitting of
  the two lobes by the ISM. The relevant driver may well be an instability
  (Opher et al.\ 2021) related to the interaction of streaming ISM neutral
  H and the ionized heliotail plasma that results in a loss of coherence of
  the heliospheric jets and leads to the mixing of ISM and heliospheric
  tail material. This instability is present in the BU model but not in the
  Moscow model (for reasons that are not clear at the present time).
  Investigation of the instability and its consequences such as
  reconnection and ISM-heliotail plasma mixing are deferred to a future
  study. Numerical solution strategies and grid resolutions might play an
  important role in the different modeling outcomes. This is a topic that
  needs to be addressed, as global heliosphere modeling suffers from the
  inability to model phenomena on comparatively small scales, including
  reconnection, the heliospheric current sheet, and instabilities.
  Therefore, additional comparisons with other models both on the scale of
  the global heliosphere and also at small scales are required to further
  address this. In future work we will use time-dependent MHD models to
  compare more realistic solutions. The results are important since the
  different heliosphere properties from these models, such as heliosheath
  thickness at high latitudes, yield different ENA signatures at different
  energy bands. These differences are potentially observable by the
  Interstellar Boundary Explorer. This will be explored in a future study.''
\end{quote}

\subsubsection{Heerikhuisen et al. (2025)}
\label{sec:heerikhuisen2025}

A significant citation from this publication is: ``In this paper we show
how numerical simulations with the same boundary conditions may evolve
differently, depending on factors such as initial conditions and run
time.''

From the abstract:

\begin{quote}
  ``Ever since the discovery of the solar wind, it has been understood that
  our solar system resides within a bubble of solar wind plasma that
  originates from the Sun. Due to its vast size, the geometry of this
  bubble, known as the heliosphere, remains a topic of conjecture and
  debate. Three-dimensional simulations, employing magnetohydrodynamics
  coupled to neutral particles, provide a key way of understanding the
  global flow dynamics on macroscopic scales. While many of these models
  suggest a comet-like shape with a single heliotail for the heliosphere,
  others have suggested geometries with two distinct outflows of solar wind
  plasma. In this paper we show how numerical simulations with the same
  boundary conditions may evolve differently, depending on factors such as
  initial conditions and run time. We find that the eventual steady state
  is the same comet-like shape, even for less than optimal choices of the
  initial condition.''
\end{quote}

From the introduction:

\begin{quote}
  ``M. Opher et al.\ (2015, 2016) presented results from a simulation that
  show a double-tailed (or `croissant-like' M. Opher et al.\ 2016)
  heliosphere with two distinct outflows. While it was suggested that this
  may herald a new picture of the heliosphere (M. Opher et al.\ 2015),
  other groups were not able to reproduce this specific geometry, even
  using the same model parameters. In fact, N. V. Pogorelov et al.\ (2015)
  showed that the assumption in M. Opher et al.\ (2015) of a unipolar
  magnetic field for the Sun significantly enhances the magnetic field in
  the heliosheath, resembling the double-tailed structure suggested by
  G. Yu (1974), though the magnetic tube structures still result in a
  single heliotail. Those authors also point out that a kinetic treatment
  of neutrals and a time-dependent SW both prevent magnetic tube structures
  from developing, and hence a single tail is the most likely geometry for
  our heliosphere. Furthermore, N. V. Pogorelov et al.\ (2017) show that
  any signature of the split-tail configuration disappears if solar cycle
  effects are taken into account. Finally, E. J. Zirnstein et al.\ (2017)
  show that a time-dependent single-tailed heliosphere reproduces the
  heliotail lobe structure observed in energetic neutral atom maps from the
  IBEX spacecraft (D. J. McComas et al.\ 2013). We also note here that
  later versions of the model first shown in M. Opher et al.\ (2015),
  namely M. Kornbleuth et al.\ (2021) and A. T. Michael et al.\ (2022), do
  not display a strong croissant-like shape. M. Kornbleuth et al.\ (2021)
  show a good agreement between their heliosphere and one obtained using
  the Moscow Model (V. V. Izmodenov \& D. B. Alexashov 2020), except for
  transition between the SW and LISM plasmas (i.e., the HP), which the
  models handle differently. The work of A. T. Michael et al.\ (2022) is
  axially symmetric, necessitating a single heliotail, but the results
  compare well to other heliospheres (D. Alexashov \& V. Izmodenov 2005;
  J. Heerikhuisen et al.\ 2006).

  ``In this paper we show that a steady-state heliospheric simulation can
  transition from a single-tailed to a double-tailed structure by changing
  the initialization method, while keeping the boundary conditions the
  same. The resulting heliosphere resembles the two-tailed structure shown
  in M. Opher et al.\ (2015) after plasma--neutral coupling is initiated,
  though all our simulations eventually reach a single-tailed steady state.
  More generally, we show how a steady-state heliosphere can be obtained
  using variations on the commonly applied time marching approach, and that
  certain strategies can speed up the convergence of the solution.''
\end{quote}

\section{Issues concerning instabilities in the heliosheath, constraints
  related to solar wind speed, the role of turbulence and the Interstellar
  Mapping and Acceleration Probe (IMAP) mission}
\label{sec:issues}

\subsection{Ma et al. (2025a)}
\label{sec:ma2025a}

The authors of the work \emph{Instabilities along the Axis of the
Heliospheric Jets} \citep{ma2025a} describe the formation of a low-speed
region in the heliosheath (HS) due to charge exchange between solar wind
and neutral hydrogen (H) and magnetic tension. This region allows
sufficient time for the instabilities to develop and induces a shear flow
inside the HS, leading to the development of a Kelvin--Helmholtz (K-H)
instability. Understanding the development of the instabilities in the HS
is crucial for comprehending the dynamic processes within the HS and the
structure of the heliosphere.

\subsection{Onubogu et al. (2025)}
\label{sec:onubogu2025}

The authors of the paper \emph{Exploring the Complex Heliotail Boundary by
an Extended Level Set Approach} \citep{onubogu2025} note that in earlier
studies the heliopause was defined using magnetohydrodynamic variables,
which provide only an approximation for capturing its position in all
directions. In the present heliosphere model, a level set method is used to
constrain the location of the heliopause (HP). The level set function,
$f_{\rm HP}$, is defined by boundary conditions, having a value of $+1$ at
the inner boundary and $-1$ at the outer boundary. The authors implemented
a level set method and simulated the heliosphere and ISM interaction to
characterize the regions of the heliosphere, specifically the HP. Their
results show that different level set criteria will produce notably
different shapes for the heliotail. In future work they would like to use
their tools to predict the shape and locations of structures in the
heliosphere in a time-dependent simulation.

\subsection{Ma et al. (2025b)}
\label{sec:ma2025b}

The authors of the paper \emph{The Coupling and Evolution of
Kelvin--Helmholtz and Rayleigh--Taylor Instabilities in the Heliosheath}
\citep{ma2025b} note that the shape and structure of the heliosphere remain
subjects of ongoing debate, with current models differing on how far the
heliospheric jets that form the two-lobe structure extend down the
heliotail and whether the surrounding interstellar medium can penetrate the
region between the lobes. Rayleigh--Taylor (RT) and Kelvin--Helmholtz (KH)
instabilities along the axis of the heliospheric jets have been proposed as
key drivers of turbulence in the heliosheath (HS). In this work, results
are presented from 225~yr MHD simulations that reveal periodic variations
in solar wind speed and magnetic field within the HS, coinciding with the
cyclic growth of RT and KH instabilities. These findings suggest that RT
and KH instabilities can coevolve and reoccur periodically, contributing to
turbulence generation in the HS. Their nonlinear development may play a
fundamental role in shaping the large-scale structure of the heliosphere,
particularly in the formation and evolution of the open heliospheric tail.

\subsection{Reisenfeld et al. (2026)}
\label{sec:reisenfeld2026}

The article \emph{Exploring the Outer Heliosphere Through ENA Observations
from IMAP} \citep{reisenfeld2026} opens in a fascinating way and contains a
wealth of knowledge. The Interstellar Mapping and Acceleration Probe (IMAP)
mission is poised to revolutionize our understanding of two of the most
central issues in heliophysics today: the nature of the interaction of the
solar wind with the local interstellar medium (LISM), and the acceleration
of energetic particles throughout the heliosphere. Here the authors focus
on the anticipated contributions to this endeavour by measurement of
hydrogen energetic neutral atoms (ENAs) that emanate from the solar
wind--LISM interface region. ENA imaging allows us to remotely sense the
structure and plasma characteristics of the heliosheath and the LISM just
beyond the heliopause, and how these regions evolve in response to solar
activity. In this paper, the authors present the IMAP mission science
specific to the study of the heliosphere/LISM interaction. They review what
they have learned about the heliosphere interaction based on prior
observations and theoretical studies. They then present the outstanding
science questions that have consequently arisen and how IMAP plans to
resolve them. In particular, in their Sect.~2.2.2, the authors discuss the
use of ENA measurements from the IMAP mission to determine the shape and
scale of the heliosphere.

It is therefore extremely interesting to see what answer the IMAP mission
will provide regarding the shape of the heliosphere.

\section{Conclusions}
\label{sec:conclusions}

Pending that answer, we conclude that it remains uncertain which model
accurately reflects the actual heliosphere. Nature calls for a theoretical
model of greater precision. In our view, the interaction between the solar
wind and the interstellar medium must be described using a system of
Boltzmann equations incorporating artificial intelligence. This conclusion
is of critical importance in the context of the planned space mission
\citep{dialynas2023}, which aims to send an interstellar probe to a
distance of several hundred astronomical units from the Sun.

\begin{acknowledgements}
  The author wishes to thank Piotr Bladek for his assistance.
\end{acknowledgements}

\end{document}